\documentclass[useAMS,usenatbib]{mn2e}
\usepackage{graphicx}
\usepackage{float}
\usepackage[usenames,dvipsnames]{xcolor}

\title[{\em NuSTAR} and {\em XMM-Newton} time lags]{Iron K and Compton hump reverberation in SWIFT~J2127.4+5654 and NGC~1365 revealed by {\em NuSTAR} and {\em XMM-Newton}}
\author[Kara et al.]{E. Kara$^{1}$\thanks{E-mail:
ekara@ast.cam.ac.uk}, A. Zoghbi$^{2,3}$, A. Marinucci$^{4}$, D. J. Walton$^{5}$, A. C. Fabian$^{1}$, 
\newauthor
G. Risaliti$^{6,7}$, S. E. Boggs$^{8}$, F. E. Christensen$^{9}$, F. Fuerst$^{5}$, C. J. Hailey$^{10}$,
\newauthor
F. A. Harrison$^{5}$, G. Matt$^{4}$, M. L. Parker$^{1}$, C. S. Reynolds$^{2}$, D. Stern$^{11}$, 
\newauthor
and W. W. Zhang$^{12}$ \\
$^{1}$Institute of Astronomy, The University of Cambridge, Madingley Road, Cambridge, CB3 OHA\\
$^{2}$Department of Astronomy, University of Maryland, College Park, MD 20742-2421, USA\\
$^{3}$Joint Space-Science Institute (JSI), College Park, MD 20742-2421, USA\\
$^{4}$Dipartimento di Matematica e Fisica, Universit\`a degli Studi Roma Tre, via della Vasca Navale 84, 00146 Roma, Italy\\
$^{5}$Cahill Center for Astronomy and Astrophysics, California Institute of Technology, Pasadena, CA 91125, USA\\
$^{6}$Harvard-Smithsonian Center for Astrophysics, 60 Garden Street, Cambridge, MA, USA\\
$^{7}$INAF Ð Osservatorio Astrofisico di Arcetri, L.go E. Fermi 5, I-50125 Firenze, Italy\\
$^{8}$Space Science Laboratory, University of California, Berkeley, CA 94720, USA\\
$^{9}$DTU Space National Space Institute, Technical University of Denmark, Elektrovej 327, DK-2800 Lyngby, Denmark\\
$^{10}$Columbia Astrophysics Laboratory, Columbia University, New York, NY 10027, USA\\
$^{11}$Jet Propulsion Laboratory, California Institute of Technology, Pasadena, CA 91109, USA\\
$^{12}$NASA Goddard Space Flight Center, Greenbelt, MD 20771, USA\\
}

\begin{document}

\date{Accepted 2014 October 13.  Received 2014 October 13; in original form 2014 May 16}

\pagerange{\pageref{firstpage}--\pageref{lastpage}} \pubyear{2014}

\maketitle

\label{firstpage}

\begin{abstract}
In the past five years, a flurry of X-ray reverberation lag measurements of accreting supermassive black holes have been made using the {\em XMM-Newton} telescope in the 0.3--10~keV energy range. In this work, we use the {\em NuSTAR} telescope to extend the lag analysis up to higher energies for two Seyfert galaxies, SWIFT J2127.4+5654 and NGC~1365.  X-ray reverberation lags are due to the light travel time delays between the direct continuum emission and the reprocessed emission from the inner radii of an ionised accretion disc.  {\em XMM-Newton} has been particularly adept at measuring the lag associated with the broad Fe K emission line, where the gravitationally redshifted wing of the line is observed to respond before the line centroid at 6.4~keV, produced at larger radii.  Now we use {\em NuSTAR} to probe the lag at higher energies, where the spectrum shows clear evidence for Compton reflection, known as the Compton `hump'.  The {\em XMM-Newton} data show Fe K lags in both SWIFT~J2127.4+5654 and NGC~1365. The {\em NuSTAR} data provide independent confirmation of these Fe K lags, and also show evidence for the corresponding Compton hump lags, especially in SWIFT~J2127.4+5654.  These broadband lag measurements confirm that the Compton hump and Fe K lag are produced at small radii. At low-frequencies in NGC~1365, where the spectrum shows evidence for eclipsing clouds in the line of sight, we find a clear negative (not positive) lag from 2--10~keV, which can be understood as the decrease in column density from a neutral eclipsing cloud moving out of our line of sight during the observation.  

\end{abstract}


\section{Introduction}
\label{intro}

\begin{table*}
\centering
\begin{tabular}{c|c|c|c|c|c}
\hline
Object & {\em NuSTAR} Obs. ID & {\em XMM} Obs. ID & Obs. Date & {\em NuSTAR} Exposure (s)& {\em XMM} Exposure (s) \\
\hline
SWIFT~J2127.4+5654& 60001110002/3 & 0693781701 & Nov 2012& 77000 & 94000  \\
& 60001110005&0693781801 & Nov 2012 & 74000& 94000  \\
& 60001110007&0693781901 & Nov 2012 & 42000& 50000  \\
NGC~1365 &60002046002/3 & 0692840201 & July 2012 & 77000 & 110000\\
&60002046005 & 0692840301 & Dec 2012 & 66000& 93000\\
&60002046007 & 0692840401 & Jan 2013 & 74000& 90000  \\
&60002046009 &0692840501& Feb 2013 & 70000 & 103000  \\
\hline
\end{tabular}
\caption{The {\em NuSTAR} and {\em XMM-Newton} observations used in this analysis. Columns show the source name, the observation ID for both telescopes, the start date, and duration of the observation for both telescopes.}
\label{obs}
\end{table*}

The X-ray emission around accreting supermassive black holes is very bright and highly variable on timescales of hours to days \citep{mchardy88}.  X-ray timing analysis can therefore be a very powerful tool for probing the innermost regions of Active Galactic Nuclei (AGN).  In this work, we study the X-ray reverberation time delays, which measure the size scales of the innermost regions in physical units, and not just gravitational units (i.e. in kilometres rather than in gravitational radii).  Reverberation, can therefore, in principal provide independent estimates of the black hole mass and spin.

\begin{figure*}
\includegraphics[width=\textwidth]{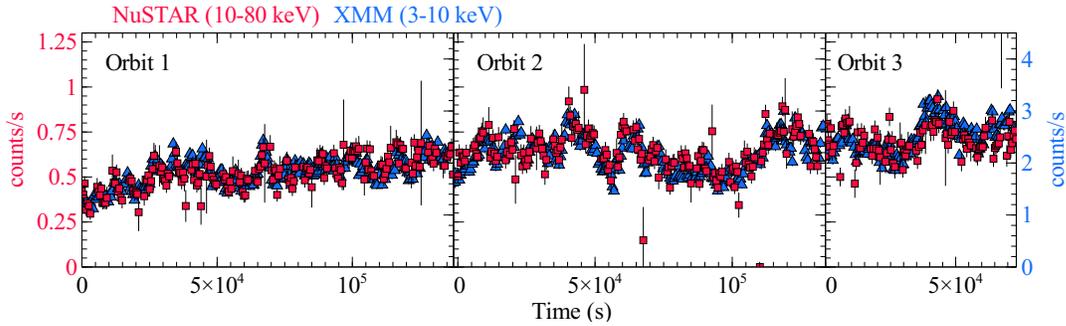}
\caption{The {\em NuSTAR} light curve from 10--80 keV in red squares overplotting the {\em XMM-Newton} light curve from 3--10~keV in blue triangles for SWIFT~J2127.4+5654.  The left y-axis refers to the {\em NuSTAR} counts, while the right y-axis refers to the {\em XMM-Newton} counts.  The {\em NuSTAR} and {\em XMM-Newton} light curves track each other closely in all three observations.}
\label{2127_lc}
\end{figure*}

\begin{figure*}
\includegraphics[width=\textwidth]{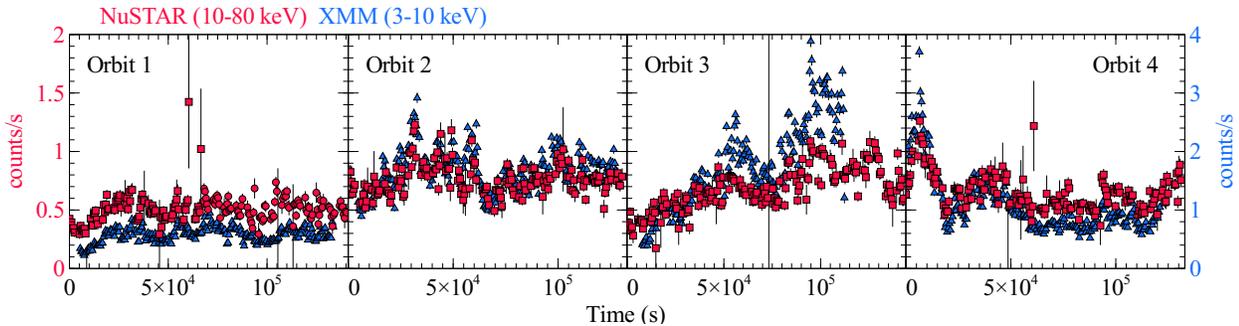}
\caption{The {\em NuSTAR} light curve from 10--80 keV in red squares overplotting the {\em XMM-Newton} light curve from 3--10~keV in blue triangles for NGC~1365.  The left y-axis refers to the {\em NuSTAR} counts, while the right y-axis refers to the {\em XMM-Newton} counts.  For this lag analysis, we focus on the third orbit where the flux and variability power are the greatest. We also present the results from the second observation, which shows some variability.}
\label{1365_lc}
\end{figure*}

X-ray reverberation is due to the light crossing time of photons around the accreting black hole.  The first robust discovery was made by \citet{fabian09} for the Narrow-line Seyfert I galaxy, 1H0707-495. In that work, the authors measured that the high-frequency variability in the reflection-dominated soft band (0.3--1~keV) was delayed with respect to the continuum-dominated hard band (1--4~keV) by 30~s.  This was interpreted as the light travel distance between the X-ray emitting corona, and the reprocessed emission off the inner accretion disc where strong gravity effects are important.  Later findings of this high-frequency `soft' lag in other NLS1s confirmed the detection \citep{emm11,demarco11,zoghbi11b,cackett13,demarco13,emm14}. Recent work by \citet{cackett14} and later by \citet{emm14} model this time lag using general-relativistic ray-tracing simulations \citep{reynolds99, dovciak04}, and compute that these lags come from small radii within $10~r_{\mathrm{g}}$ of the central supermassive black hole.  At low-frequencies, a separate process dominates the lags, and instead of a soft band lag, there is a hard band lag. This low-frequency hard lag has been observed in black hole binaries \citep{miyamoto89, nowak99} and AGN \citep{papadakis01,arevalo06b,mchardy07}. While its origin is not well understood, the prevailing interpretation is that the lags are due to mass accretion rate fluctuations in the disc that propagate inwards and are transferred up to the corona, causing the soft continuum emission from large radii to respond before the hard emission at small radii \citep{kotov01,arevalo06}.

While the high-frequency soft lag was detected in a number of sources, the interpretation was still a contentious issue.  \citet{miller10} and later \citet{legg12} suggested an alternative interpretation for the soft lags.  In these papers, the authors interpret the high-frequency soft lag as a mathematical oscillatory effect from low-frequency, distant reverberation off a reprocessing medium at $100-1000~r_{\mathrm{g}}$.  The low-frequency, hard lag, is then understood as the long-timescale reverberation lag.  The frequency-dependence of the soft lag could be similarly well modelled with a distant reprocessor.

However, the discovery of the high-frequency iron K reverberation lag has put further constraints on physical models, confirming the relativistic reflection interpretation.  \citet{zoghbi12} first discovered the iron K reverberation lag in the bright Seyfert galaxy, NGC~4151.  In that work, the authors computed the high-frequency lag-energy spectrum and found that the continuum emission varies first, followed by a delayed response in the red wing of the line, and lastly by the line centroid.  This is naturally understood as the reverberation off the inner accretion disc, where there is a smaller light travel time between variations of the corona and the innermost radii (where the red wing is produced), than between the corona and the larger radii of the accretion disc (where the line centroid originates).  Since this initial discovery, the high-frequency iron K lag has been found in a number of Seyfert galaxies \citep{kara13a,kara13b,zoghbi13a}.  It has been shown that while there are strong signatures of reflection in the high-frequency lags, the low-frequency lags are featureless, and are not clearly associated with any reflection \citep{kara13c}.  In one source, NGC~6814, where the spectrum is well described with only an absorbed powerlaw and minimal neutral reflection, there is still a clear low-frequency, hard lag \citep{walton13}, which further indicates that this low-frequency hard lag is {\em not} associated with reflection, as proposed by \citet{miller10}.  Lastly, detailed GR ray-tracing models by \citet{reynolds99} have also been shown to describe the high-frequency lags well \citep{cackett14}.  

The next frontier for reverberation lag studies is to extend the lag-energy spectrum up to high energies above 10~keV, where the reflected emission peaks. This is now possible with the {\em Nuclear Spectroscopic Telescope Array} \citep[{\em NuSTAR}; ][]{harrison13}, the first high-energy focusing X-ray telescope in orbit.  {\em NuSTAR} is 100 times more sensitive than previous instruments probing the 10-80~keV range, and in the past two years has made significant contributions in measuring the high-energy spectra of AGN.  This started with Seyfert galaxy NGC~1365, where the reflection feature of the Compton hump was clearly detected \citep{risaliti13}. Since then, the Compton hump has been confirmed in a number of other objects that contain broad iron lines, including MCG-6-30-15 \citep{marinucci14b}, and Mrk 335 \citep{parker14c}.  Now that the Compton hump has been confirmed in the energy spectrum, we want to search for the feature in the lags.  \citet{zoghbi14} presented the first analysis of high-frequency time lags above 10~keV, and we expand on this work with an analysis of SWIFT~J2127.4+5654 and NGC~1365.

SWIFT J2127.4+5654 ($z=0.0144$) is a NLS1 galaxy that was first detected with the {\em Swift}/BAT in the 15--150~keV band \citep{tueller05}.  The source was observed in 2007 with {\em Suzaku}-XIS for 92~ks \citep{miniutti09}. The authors detected a broad Fe~K emission line, which they used to infer a black hole spin of $a=0.6\pm0.2$. The result was confirmed by \citet{patrick11}, and later by \citet{sanfrutos13} using a 130~ks {\em XMM-Newton} observation.  \citet{marinucci14} presented the 300~ks joint {\em XMM-Newton}/{\em NuSTAR} observation of this source, confirming the broadened Fe~K emission line and discovering a clear Compton hump.  Also in that work, we analysed the {\em XMM-Newton} data and found a high-frequency Fe~K reverberation lag that was not as broad spectrally as those found in sources with maximally spinning black holes. This reverberation result independently confirmed a narrower relativistically broadened iron line and a compact X-ray source in SWIFT~J2127.4+5654.

NGC~1365 ($z=0.0055$) is a Seyfert 1.9 galaxy that also shows strong evidence for a relativistically broadened iron line, this time implying a maximally spinning black hole \citep{risaliti09,walton10, brenneman13}.  In addition, it is known to have complex and variable absorption, with evidence for a warm absorber and even cold, eclipsing material along the line of sight in some observations \citep{risaliti05a,risaliti05b,maiolino10}.  Recently, the source was observed for four {\em XMM-Newton} orbits in the joint {\em XMM-Newton}/{\em NuSTAR} AGN campaign. The four observations show remarkable variability, and \citet{walton14} has recently shown that this variability can largely be explained by absorption variability, and that underlying the complex absorption structure, the relativistically broadened iron line and Compton hump are always present.  Orbit 3 of this observation was the most unobscured of the 4 observations.  Principal Component Analysis of the {\em XMM-Newton} observations also showed that Orbit 3 has the most intrinsic source variability, though some absorption variability is still present \citep{parker14b}.  For this reason, we focus our time lag analysis on Orbit 3, but show results from the other more absorbed orbits, which yield no evidence of Fe~K reverberation lags.

This paper is structured as follows: In Section~\ref{obs_sec} we describe the two types of lag analysis used in this work, the standard Fourier technique and the more recently developed maximum-likelihood technique for unevenly sampled light curves \citep{zoghbi13}.  We present our results for SWIFT J2127.4+5654 and NGC~1365 in Section~\ref{results} and discuss them in Section~\ref{discuss}.

\section{Observations and Data Analysis}
\label{obs_sec}

\begin{figure*}
\includegraphics[width=\textwidth]{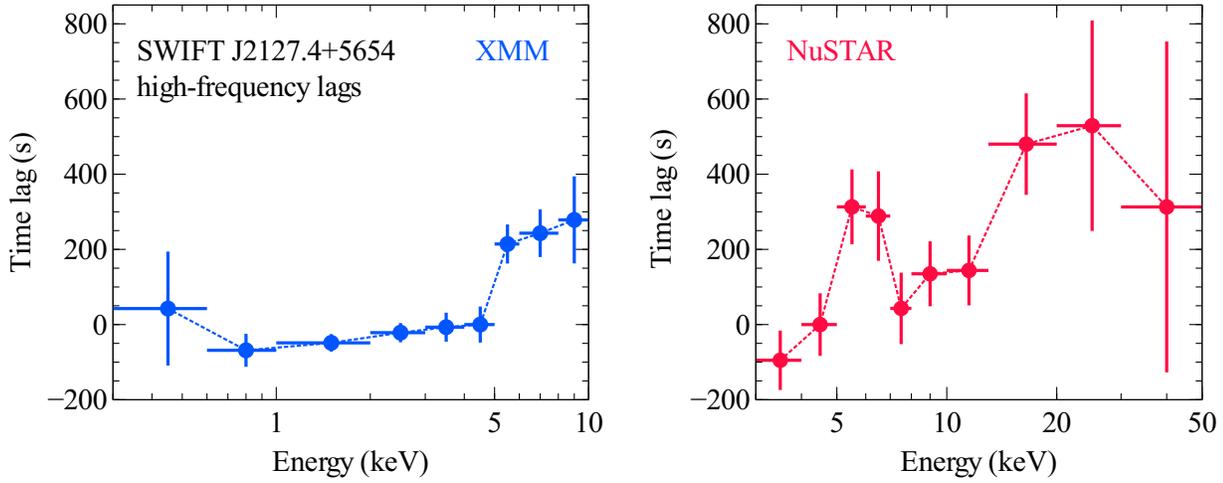}
\caption{The high-frequency lag-energy spectra for SWIFT J2127.4+5654 using {\em XMM-Newton} (left) and {\em NuSTAR} (right). The lag is calculated in the frequency range, $[0.4-4.5] \times 10^{-4}$~Hz.  The {\em XMM-Newton} lag-energy spectrum shows a sharp increase above 5~keV. The {\em NuSTAR} lag shows the same peak at 5--7~keV, and another peak at $\sim 20$~keV, the energy of the Compton hump.  To make the comparison easier to see, both the {\em XMM-Newton} and {\em NuSTAR} lags have been scaled so that the lag at 4--5~keV is zero. Due to lower statistics at the highest energies in {\em XMM-Newton}, we cannot disentangle the lag at the blue wing of the line from the start of the rise of the Compton hump, as is evident in {\em NuSTAR}.  However, the lag results between the two instruments are consistent within error.  See Fig.~\ref{model} for further comparison of the {\em XMM-Newton} and {\em NuSTAR} lag-energy spectrum.}
\label{2127_high_lagen}
\end{figure*}

\begin{figure*}
\includegraphics[width=\textwidth]{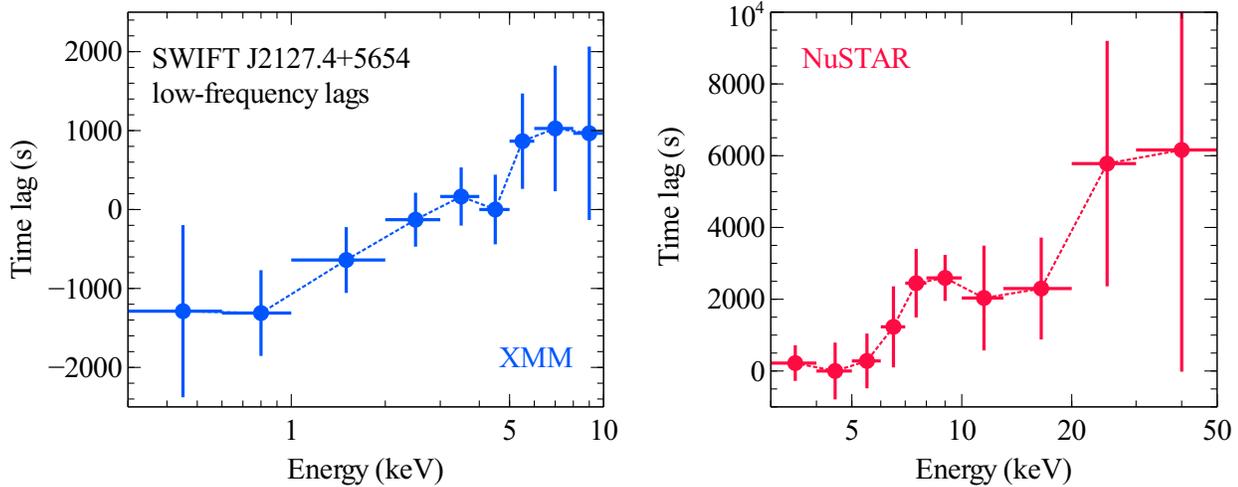}
\caption{The low-frequency lag-energy spectra for SWIFT J2127.4+5654 using {\em XMM-Newton} (left) and {\em NuSTAR} (right).  Note the different y-axis scales between the two figures, and between these figures and those in Fig.~\ref{2127_high_lagen}. The lag is calculated in the frequency range less than $0.4 \times 10^{-4}$~Hz.  The lag appears to increase with energy in both the {\em XMM-Newton} and {\em NuSTAR} bands, though the {\em NuSTAR} low-frequency lag shows a peak at 7--8~keV.  Again, the {\em XMM-Newton} and {\em NuSTAR} lags have been scaled so that the lag at 4--5~keV is zero.}
\label{2127_low_lagen}
\end{figure*}

\subsection{Data Reduction}

All of the data used in this work come from the joint {\em XMM-Newton} and {\em NuSTAR} AGN campaign, where observations were taken simultaneously (Table~\ref{obs}).  SWIFT J2127.4+5654 was observed with both instruments for 300~ks, and NGC~1365 was observed for 500~ks.  We use data from the {\em XMM-Newton} observatory \citep{jansen01} taken with the EPIC-PN camera \citep{struder01}. A detailed description of the {\em XMM-Newton} and {\em NuSTAR} data reduction for SWIFT J2127.4+5654 can be found in \citet{marinucci14}, and for NGC~1365 in \citet{walton14}.   

\subsection{Time lag measurements}

The X-ray signals from accreting black holes are highly variable (as seen in Figs.~\ref{2127_lc} and \ref{1365_lc}, which show the light curves of SWIFT~J2127.4+5654 and NGC~1365 in {\em XMM-Newton} and {\em NuSTAR}). In AGN, the variability is observed on a range of timescales from hours to days, and this fact allows us to measure time delays between light curves of different energies.  There are several approaches to measuring these time delays, depending on the timescale one wants to probe and the type of time signal available.

In this section we describe the two different techniques implemented here for measuring time lags: the Fourier technique and the maximum-likelihood technique.  {\em XMM-Newton} is in an elliptical orbit with a 48-hour orbital period. This gives long, uninterrupted exposures, which are ideal for the traditional Fourier techniques, where the lowest frequency probed is the orbital frequency (1/orbital period).  This technique is not possible with {\em NuSTAR} because the telescope is on a low Earth orbit, with an orbital period of $\sim 90$ minutes.  For AGN, we typically want to probe frequencies lower than the {\em NuSTAR} orbital frequency, and therefore we employ the maximum-likelihood technique, which accounts for the orbital gaps in the data.  

\subsubsection{The Fourier technique}

For the Fourier technique, we follow the methodology outlined in \citet{nowak99}, and explained further in \citet{vaughan03} and \citet{uttley14}.  We produce light curves in different energy bands in 10~s bins.  We take the discrete Fourier transform of each light curve, which can be expressed in its phasor form as the product of its amplitude and complex exponential phase.  Taking the complex conjugate reverses the sign of the phase.  To calculate the phase difference between two light curves, we multiply the Fourier transform of one light curve by the complex conjugate of the Fourier transform of another.  This product is known as the cross spectrum, and its phase is simply the phase difference between the two light curves. The cross spectrum is averaged in frequency bins, and this frequency-dependent phase can be converted into a frequency-dependent time lag by dividing by $2 \pi f$ where $f$ is the middle-frequency of the logarithmic bin.  Throughout this paper, a positive lag is defined such that the hard band light curve is delayed with respect to the soft band.  The 1-$\sigma$ errors are determined following Equation 16 of \citet{nowak99}, and are based on the number of frequency bins sampled and the coherence between the light curves.

\subsubsection{The maximum-likelihood technique}

The maximum-likelihood technique for parametric model fitting by estimating the covariance matrix has been discussed in several works \citep{dempster72,anderson73,stein86}.  It was applied to parameter estimation for sparse 2D power spectra of the cosmic microwave background \citep{bond98}, and then later to X-ray light curves in \citet{miller10}. The technique was developed by \citet{zoghbi13} for the application of measuring X-ray time lags, and has been shown through Monte Carlo simulations to give the same results as the standard Fourier techniques.

The method fits for the most likely variability powers and time lags given the observed data. The technique relies on the fact that the autocorrelation is the Fourier transform of the Power Spectral Density (PSD).  If we model the PSD as a step function (parameterized by the power in each pre-defined frequency bin), then we can compute the maximum likelihood between the observed autocorrelation and the model PSD parameters. Analogously, the cross correlation is the Fourier transform of the cross spectrum, and in this case the model parameters that we compute are the amplitude and phase of the cross spectrum. Similar to the standard Fourier techniques, that phase lag is then converted into a time lag by dividing by $2 \pi f$.  The errors presented are computed by stepping through the parameters in the likelihood function and taking the 68\% uncertainty as the value that changes $-2$log($\mathcal{L}/\mathcal{L}_{\mathrm{max}}$) by 1 \citep{zoghbi13}.

As the autocorrelation and cross correlation matrices can be computed with unevenly sampled light curves, this technique is the best way to find the low-frequency time lags in light curves from telescopes in a low Earth orbit, such as {\em Suzaku} and {\em NuSTAR}. For these telescopes, we are in the regime where the light curves have a constant time bin width, but there are gaps of missing data due to Earth occultation.  We direct the reader to \citet{zoghbi13} and Appendix~A in this paper for more details on the maximum-likelihood method. 

\section{Results}
\label{results}

\subsection{SWIFT J2127.4+5654}
\label{2127}

We presented the first discovery of reverberation lags in SWIFT~J2127.4+5654 in \citet{marinucci14}, using these {\em XMM-Newton} observations. In that paper, we looked at the lag-frequency spectra between the soft (0.3--1~keV) and mid (1--5~keV) bands, and found that the mid-band lags the soft band at frequencies below $\sim 3 \times 10^{-5}$~Hz.  Considering the lag-frequency spectrum between the 3--5~keV and 5--8~keV bands revealed a further high frequency, hard band lag at frequencies below $\sim 3 \times 10^{-4}$~Hz.  We used these lag-frequency spectra to decide which frequencies to probe for the lag-energy spectra. We direct the interested reader to \citet{marinucci14} for a detailed discussion of the lag-frequency spectrum, and we simply highlight here some of the main reverberation results to compare with the higher energy {\em NuSTAR} results.

Fig.~\ref{2127_high_lagen} shows the high-frequency lag-energy spectra for SWIFT~J2127.4+5654 using {\em XMM-Newton} in the left panel and {\em NuSTAR} in the right panel.  The {\em XMM-Newton} and {\em NuSTAR} lag-energy spectra are both computed by measuring the lag in each energy channel of interest with respect to a broad reference band.  The reference band is chosen to be the entire energy band (0.3--10~keV for {\em XMM-Newton} and 3--50~keV for {\em NuSTAR}), with the channel of interest removed so that the noise is not correlated. This choice of reference band does not affect the relative shape of the lag-energy spectrum (See Appendix~B for more details).  The lag is read from bottom to top, i.e. the smaller the lag, the earlier the signal arrived at the detector.  It is important to note that the meaningful quantity is the relative lags between energy bins.

The {\em XMM-Newton} lag-energy spectrum on the left shows little or no lag between energies from 0.3--5~keV, but above 5~keV the emission is delayed by $\sim 250$~s.  In \citet{marinucci14}, we interpreted this sharp increase at $>5$~keV as the reverberation lag from a relativistically broadened Fe~K emission line.  What is particularly interesting about this result is that SWIFT~J2127.4+5654 has been suggested to have a spin of $a=0.56$, using the iron line fitting method, implying that the ISCO is larger than for maximally spinning black holes. The reverberation result shows in a unique and model-independent way that the Fe K line is not as broad in this source as in other sources where maximally spinning black holes have been inferred \citep[e.g. NGC~4151, 1H0707-495, IRAS~13224-3809; see Fig.~6 of][]{kara14}.  

There is little lag associated with the soft excess below 1~keV.  The reflection fraction below 1~keV is low in this source, and therefore, we do not expect a bigger soft lag than what is observed.  

\begin{figure*}
\includegraphics[width=\textwidth]{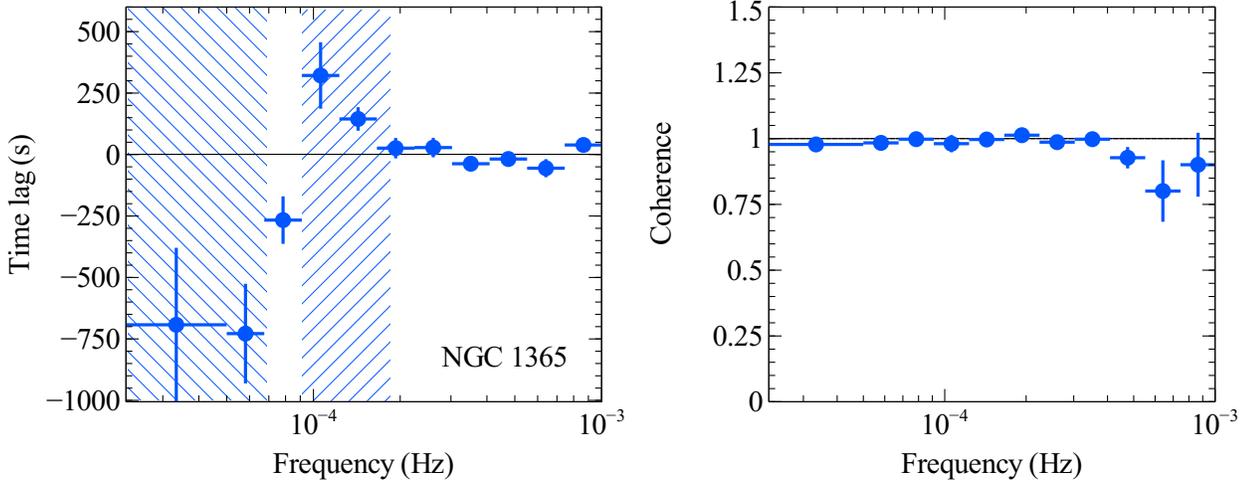}
\caption{({\em left:})  The lag vs temporal frequency for orbit 3 of NGC~1365 between the 2--4 keV and 4--7~keV bands.  The positive lag from $[0.9-2] \times 10^{-4}$~Hz shows that the hard band lags behind the soft band. This is the behaviour expected from Fe K reverberation, so we will look in this frequency range (as shown in the rightmost shaded box) for the high-frequency lag-energy spectrum.  We usually see the positive lag continue to very low frequencies, but in the case of NGC~1365, the lag becomes negative below $\sim 9 \times 10^{-5}$~Hz.  We will explore this low-frequency regime (indicated by the shaded box on left) with the lag-energy spectrum.  ({\em right:}) The coherence as a function of frequency between the same 2--4~keV and 4--7~keV bands.  The coherence is high at all frequencies we explore, and only drops off at high frequencies where Poisson noise starts to dominate the power spectrum.  Even at low frequencies, where we see the lag switch from positive to negative, we find the coherence is high. Therefore, we can have confidence in our measurement of the lag at these frequencies.}
\label{xmm_lag_freq}
\end{figure*}

The panel of the right shows the corresponding lag using the {\em NuSTAR} data at the same frequency. Again we see the sharp increase at 5--7~keV that decreases above 7~keV. The fact that the {\em XMM-Newton} lag-energy spectrum does not show the blue wing of the line may be because the 8--10~keV band is probing the lag associated with the Compton hump, which is clear in the {\em NuSTAR} lag-energy spectrum (see the left panel of Fig.~\ref{model} for further comparison of the {\em NuSTAR} and {\em XMM-Newton} lag-energy spectra).  {\em NuSTAR} allows us to now clearly determine the blue wing of the Fe~K line above 7~keV. Above 10~keV, the lag increases again, at the energy of the Compton hump.  The amplitude of the Fe~K lag in the {\em XMM-Newton} data is roughly 250~s, while the amplitude of the Fe~K lag in the {\em NuSTAR} data is roughly 300~s. These amplitudes are consistent within these 68.3\% error bars. We will discuss the amplitudes of the lag, and their interpretation as light travel time delays further in the discussion (Section~\ref{discuss}).

At low frequencies (Fig.~\ref{2127_low_lagen}), we see the lag increases with energy in both the {\em XMM-Newton} (left) and {\em NuSTAR} data (right).  In the {\em XMM-Newton} band from 0.3--10~keV, the lag-energy spectrum shows fewer features than at high frequencies, and as we probe higher energies in the {\em NuSTAR} band, we find the same general increase in lag with energy, but with additional features. There is a noticeable increase in the lag at 7~keV, which also corresponds to the sharp decrease in the lag at high frequencies. The origin of this low-frequency lag is not well understood, and we will discuss these results further in Section~\ref{discuss}.

\subsection{NGC~1365}
\label{1365}

NGC~1365 shows dramatic absorption variability between orbits. \citet{walton14} showed that the first and fourth {\em XMM-Newton} orbits are highly absorbed, which causes the flux below 10~keV to be significantly attenuated (see also Rivers et al., {\em in prep}).  This strong absorption inhibits the measurement of the lag.  Orbit 3 is the least absorbed, so we focus our attention on this orbit.  For completeness, we also complete the analysis of Orbits 1, 2 and 4..  

As this is the first study of lags in this source, we will present the lag-frequency and lag-energy results from {\em XMM-Newton} alone before probing higher energies with {\em NuSTAR}.

\subsubsection{The XMM-Newton lags}

Fig.~\ref{xmm_lag_freq} shows the lag (left) between the 2--4~keV band and the 4--7~keV band.  There is a clear positive (hard) lag at frequencies $[0.9-2] \times 10^{-4}$~Hz.  A positive hard lag  at these energies can either be an indication of an Fe~K lag or a featureless continuum lag, and so further study of the lag-energy spectrum is required to understand the origin.  

At frequencies below $9 \times 10^{-5}$~Hz, the lag switches from positive to negative, indicating that on long timescales, the soft band light curve lags behind the hard band.  This behaviour is not typically seen in the lag-frequency spectrum between 2--4~keV and 4--7~keV.  Again, we will look at the low-frequency lag-energy spectrum to investigate the lag further. 

We compute the coherence between the same two energy bands to check whether a reliable measurement of the lag can be made at these frequencies.  The right panel of Fig.~\ref{xmm_lag_freq} shows the frequency-dependent coherence between 2--4~keV and 4--7~keV.  The coherence calculates to what degree one light curve is a simple linear transformation of the other \citep{vaughan97}.  A maximum coherence of 1 indicates that they are complete linear transforms of each other.  The coherence must be high (though not necessarily 1) in order to reliably measure the lag \citep{kara13a}.  The coherence between these two light curves is nearly 1 at all frequencies probed, and only begins to drop at high-frequencies where the power spectrum becomes dominated by Poisson noise.  This gives us confidence in our measurement of the lag, allowing us to move on to the lag-energy spectrum to explore the high-frequency positive lag, and the unusual negative lag at low-frequencies.

\begin{figure*}
\includegraphics[width=\textwidth]{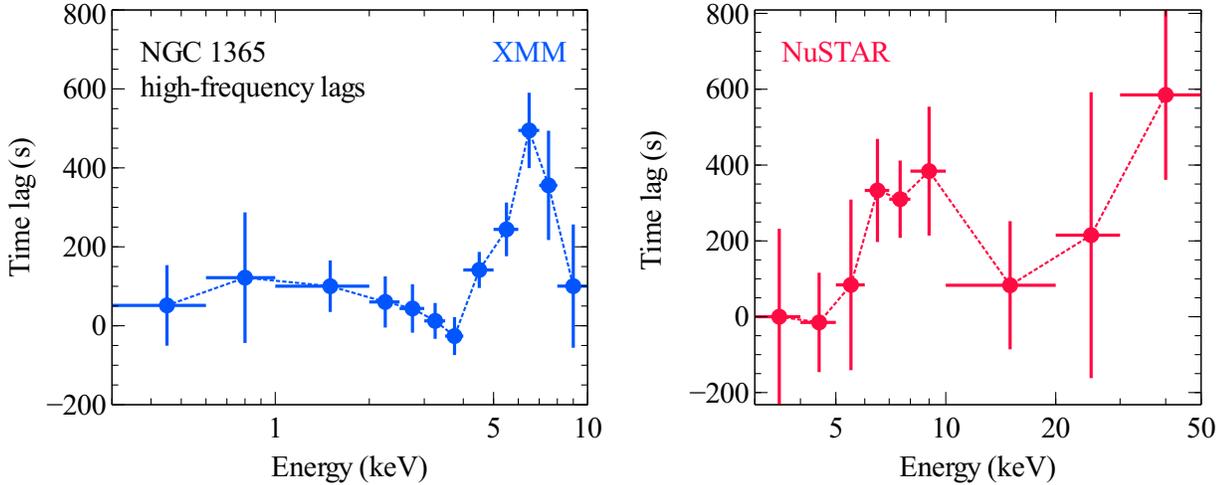}
\caption{The high-frequency lag-energy spectra for Orbit 3 of NGC~1365 using {\em XMM-Newton} (left) and {\em NuSTAR} (right). The lag is calculated in the frequency range, $[0.9-1.9] \times 10^{-4}$~Hz for both the {\em XMM-Newton} and {\em NuSTAR} data. The {\em XMM-Newton} and {\em NuSTAR} lags shows the same peak at the energy of the Fe~K line.  Again to ease comparison, both the {\em XMM-Newton} and {\em NuSTAR} lags have been scaled so that the lag at 3--4~keV is zero. }
\label{1365_high_lagen}
\end{figure*}

\begin{figure*}
\includegraphics[width=\textwidth]{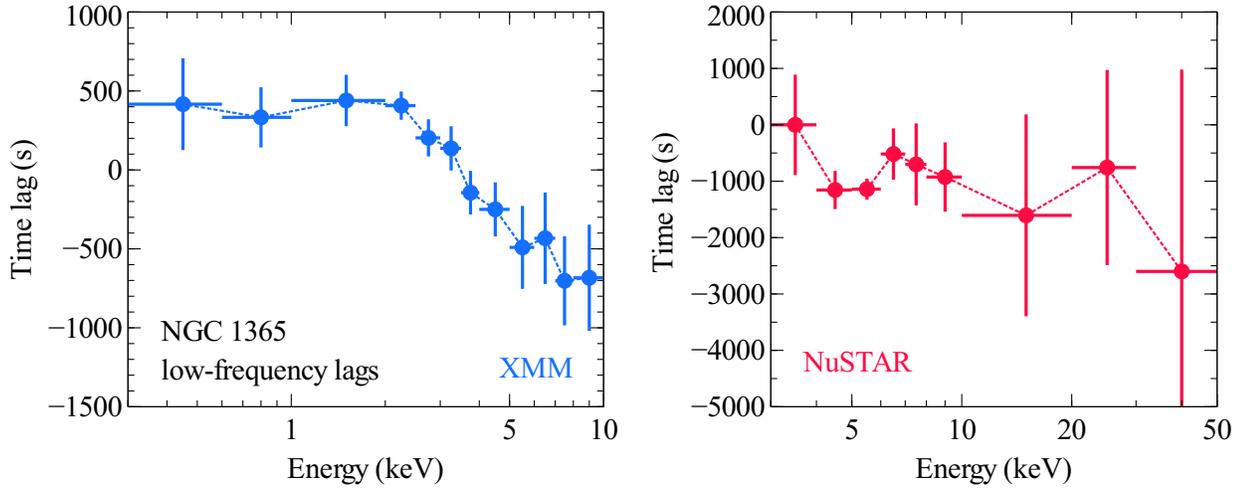}
\caption{The low-frequency lag-energy spectra for Orbit 3 of NGC~1365 using {\em XMM-Newton} (left) and {\em NuSTAR} (right).  Note the different y-axis scales between the two figures, and between these figures and those in Fig.~\ref{1365_high_lagen}.  The lag is calculated in the frequency range less than $0.7 \times 10^{-4}$~Hz.  In the {\em XMM-Newton} band, the lag drops above 2~keV, and in the {\em NuSTAR} band, the lag has large error bars, making it largely consistent with zero.  This soft lag at 2--10~keV is different from the low-frequency behaviour usually found in AGN with X-ray time lags.  Again the {\em XMM-Newton} and {\em NuSTAR} lags have been scaled so that the lag at 3--4~keV is zero.}
\label{1365_low_lagen}
\end{figure*}

\begin{figure}
\includegraphics[width=\columnwidth]{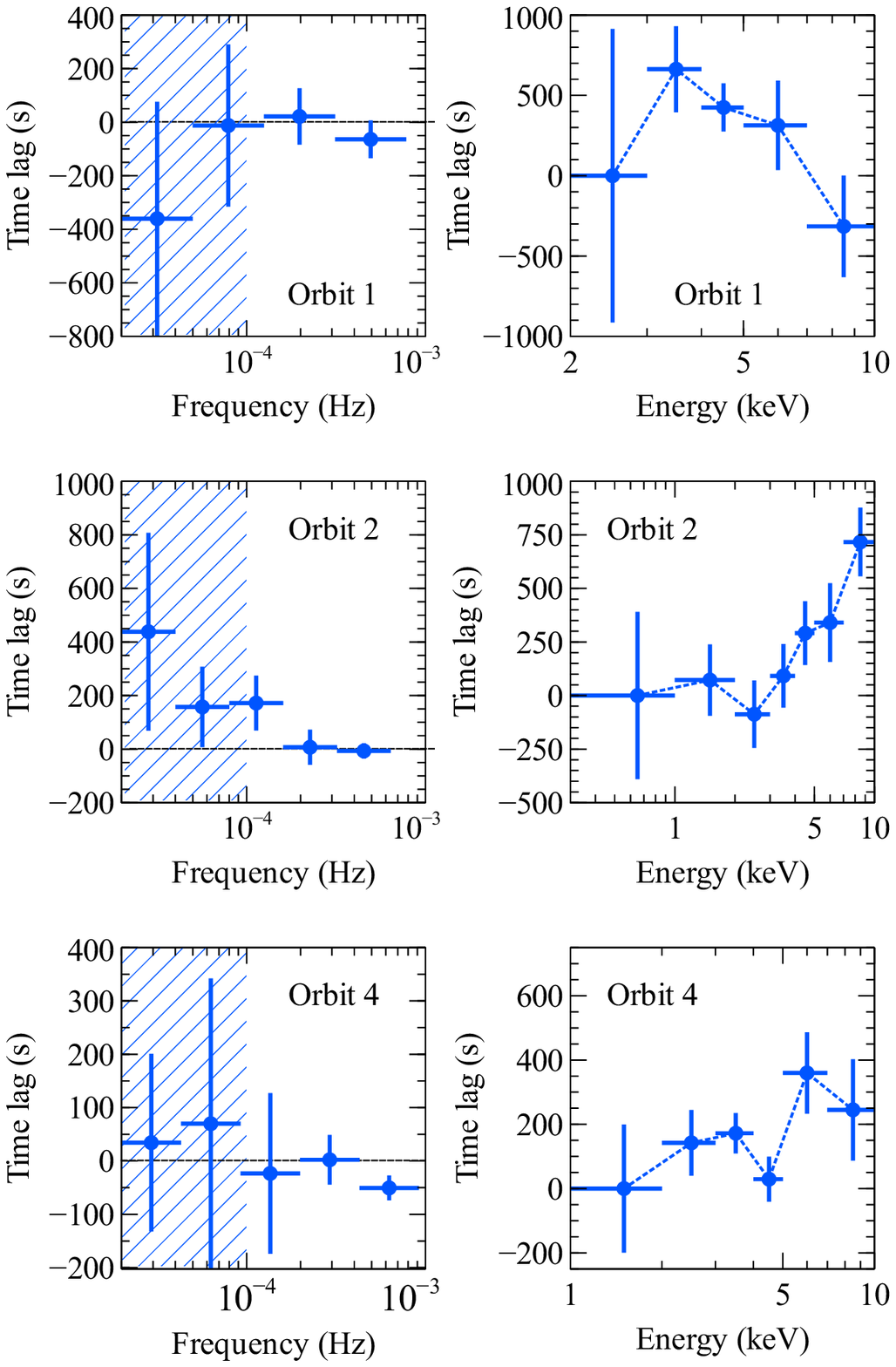}
\caption{The NGC~1365 lag-frequency spectrum ({\em left column}) and lag-energy spectrum ({\em right column}) of Orbit 1 ({\em top}), Orbit 2 ({\em middle}) and Orbit 4 ({\em bottom}). Note the different y-axis scaling between plots.  The lag vs. frequency spectra are between 2--4~keV and 4--7~keV, and the lag-energy spectra refer to frequencies below $10^{-4}$~Hz, as indicated by the shaded region in the lag-frequency spectra. The coherence (and power) are low in Orbit 1, so caution is given to these lag results.  Orbit 2 shows a hard lag at frequencies $< 10^{-4}$~Hz. The data quality are poor in the fourth orbit, though the best fit to a powerlaw model has an index of +0.1.  These results show that the low-frequency lags in NGC~1365 change from soft band leading to hard band leading between observations. }
\label{orb124}
\end{figure}

The left panel of Fig.~\ref{1365_high_lagen} shows the high-frequency lag-energy spectrum ($[0.9-1.9] \times 10^{-4}$~Hz) for Orbit 3 using {\em XMM-Newton} data alone. There is little variability power at very soft energies, as expected given the previously discovered diffuse thermal emission that dominates below 1~keV \citep{wang09}.  This lack of variability causes the error bars to be large below 1~keV even though the effective area is highest at these soft energies. At higher energies, where the variability power is high, a clear Fe~K lag is detected, with the usual `dip' in the lag at 3--4~keV, as seen in the lag-energy spectrum of 1H0707-495, IRAS~13224-3809 and several other sources with maximally spinning black holes \citep{kara13a,kara13b}. The amplitude of the Fe~K lag between 3~keV and 6~keV is roughly 500~s.

The panel on the left of Fig.~\ref{1365_low_lagen} shows the low-frequency lag-energy spectrum at frequencies below $7 \times 10^{-5}$~Hz.  The lag steadily decreases with energy above 2~keV.  There is no indication of an Fe K feature in this lag-energy spectrum.  Usually the low-frequency lag-energy spectrum increases steadily with energy (as was the case with SWIFT J2127.4+5654, and many other Seyfert galaxies), but here we see the opposite trend. 
 
\subsubsection{Lags in Orbits 1, 2 \& 4}

For completeness, we show the lags in the first, second and fourth orbits, where the variability is lower than in Orbit 3.  The coherence in Orbits 2 \& 4 is close to unity at low frequencies, but Orbit 1, which shows the greatest absorption, has very little variability power, and both the raw and noise corrected coherences are consistent with zero at the frequencies probed.  The \citet{nowak99} error estimation for the lag breaks down at low coherence \citep{vaughan97,uttley14}, and so caution is given in interpreting the lags from Orbit 1.  

Fig.~\ref{orb124} shows the {\em XMM-Newton} lags for the first, second and fourth orbits.  The left column shows the lag-frequency spectra for the same energy bins as in Fig.~\ref{xmm_lag_freq}, between 2--4~keV and 4--7~keV. The right column shows the lag vs. energy for low frequencies below $10^{-4}$~Hz.  We cannot probe energies below 2~keV for Orbit 1 and 1~keV for Orbit 4, as the variability power in the soft bands is at the level of the Poisson noise.  The first thing we notice is that none of the other orbits show the clear soft lag at low frequencies that we see in Orbit 3.  In fact, Orbit 2 shows a clear hard lag down to the lowest frequencies.  Orbit 4 shows hints of a hard lag, and the best fit powerlaw model to this lag-energy spectrum has a photon index of +0.1.  Orbit 1, which has low coherence, seems to have a negative slope from 3--10~keV.

We do not find evidence for an iron K reverberation lag in any of the other orbits.  The clearer results from Orbit 3 are to be expected as the flux and rms variability are higher during that observation, and therefore the intrinsic source continuum can be isolated.

\subsubsection{The XMM+NuSTAR lags}

Fig.~\ref{1365_high_lagen} shows the high-frequency lag-energy spectra of the third orbit for NGC~1365, comparing {\em XMM-Newton} on the left with {\em NuSTAR} on the right.  The {\em NuSTAR} frequency range ($[0.9-1.9] \times 10^{-4}$~Hz) is the same as used for the {\em XMM-Newton} data.  While the error bars are bigger using the {\em NuSTAR} data, the low-energy lag peaks at 6--8~keV, just as in the {\em XMM-Newton} data. 
The amplitude of the lag is roughly 500~s, which is within error of the {\em XMM-Newton} Fe~K amplitude.  Unfortunately, we cannot constrain much above 10~keV, though the lag does appear to increase.  The reason for the less significant lag in NGC~1365 compared to SWIFT~J2127.4+5654 is likely due to the shorter exposure in NGC~1365.  The flux and the intrinsic source variability are similar above 10~keV, but we have about one third the amount of high-quality data for NGC~1365.

Fig.~\ref{1365_low_lagen} shows the lag-energy spectra at low frequencies for {\em XMM-Newton} on the left, and {\em NuSTAR} on the right.  The {\em XMM-Newton} lags show a steady drop in the lag above 2~keV.  The {\em NuSTAR} lags cannot be well constrained, and the best fit powerlaw to the 3--50~keV band gives an index of -0.01, consistent with a straight line or zero lag.  We discuss the possible origin of this low-frequency soft lag in the next section.

The {\em NuSTAR} lag analysis was completed for the second orbit, as well, but the lag was consistent with zero at all frequencies.

\section{Discussion}
\label{discuss}

In the previous section, we presented the X-ray time lag analysis of SWIFT~J2127.4+5654 and NGC~1365 with {\em XMM-Newton} and {\em NuSTAR}.  Both sources show clear high-frequency Fe~K reverberation lags in their {\em XMM-Newton} lag-energy spectra.  SWIFT~J2127.4+5654 shows a clear Fe~K lag using {\em NuSTAR}, and, importantly, a clear lag associated with the Compton hump.  The lag associated with the Compton hump reflection was first found recently in the Seyfert galaxy MCG-5-23-16 \citep{zoghbi14}.  NGC~1365 also shows hints of these reverberation features in the {\em NuSTAR} high-frequency lag-energy spectrum, though the result is not as clear.

At low-frequencies, we find that SWIFT~J2127.4+5654 has a lag that increases with energy, similar to many other Seyfert galaxies. NCG~1365, however, shows a clear low-frequency soft lag in the {\em XMM-Newton} band, which appears to plateau above 10~keV.  In this section, we discuss the origin of both the low- and high-frequency lags in both sources.

\subsection{Interpretation of the high-frequency lags}
\label{highfreq}

High-frequency X-ray time lags are now commonly observed in variable Seyfert galaxies.  These short timescale lags are well described by reverberation of a small scale reprocessor. The iron K lag gives strong evidence that the lags are associated with reflection off the inner accretion disc. Now, the accompanying Compton hump lag confirms this interpretation.  Here, we compare the amplitude of the high-frequency lags from {\em XMM-Newton} and {\em NuSTAR}.

In order to convert the amplitude of the observed lag into the light travel time between the corona and the accretion disc, we must account for dilution effects.  We measure the lag between energy bands that contain contributions from both the continuum and the reflected emission.  This causes the amplitude of the observed lag to be smaller, or `diluted'.  The non-varying and the uncorrelated components do not contribute to the lag (i.e. neutral reflection from distant material), therefore one need only consider the dilution from the correlated, variable components (i.e. the powerlaw and ionised reflection components).  The amount of dilution can be estimated by the ratio of the reflection to powerlaw flux, referred to here as the reflection fraction.  Strictly speaking, the amount of dilution is equal to the high-frequency reflection fraction, and therefore one should measure the reflection fraction from fitting the high-frequency covariance spectrum, but the signal-to-noise of the covariance spectrum is not large enough to make meaningful constraints on these components, and so here, we assume the dilution is the reflection fraction of the mean spectrum.

\begin{figure*}
\includegraphics[width=\textwidth]{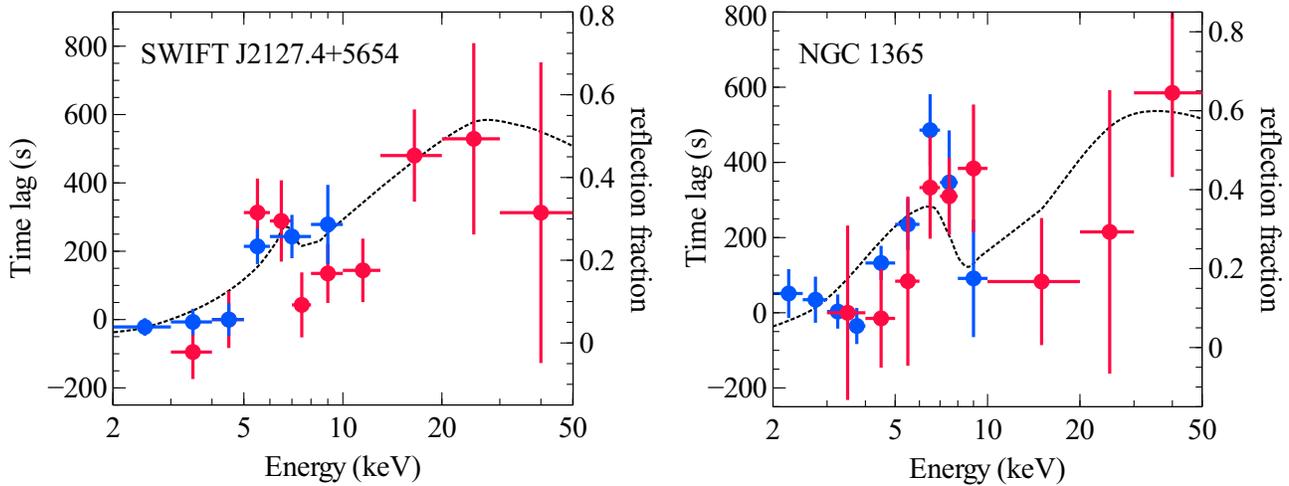}
\caption{The {\em XMM-Newton} and {\em NuSTAR} lags overplotted for SWIFT~J2127.4+5654 on the left and NGC~1365 on the right.  As before, the blue points refer to the {\em XMM-Newton} lags and the red to the {\em NuSTAR} lags. The black dotted line (which corresponds to the axis on the right) shows the reflection fraction, indicating the relative amount of dilution at a particular energy.}
\label{model}
\end{figure*}

Fig.~\ref{model} shows the {\em XMM-Newton} and {\em NuSTAR} high-frequency lag-energy spectra overplotted, for SWIFT~J2127.4+5654 on the left and NGC~1365 on the right.  Both have been scaled, so that the {\em XMM-Newton} and {\em NuSTAR} lags match at 4--5~keV for SWIFT~J2127.4+5654 and at 3--4~keV for NGC~1365.  This plot shows that the {\em XMM-Newton} and {\em NuSTAR} lag results are consistent within error. The black dotted line in both figures shows the reflection fraction (ionised reflection flux divided by the total powerlaw plus reflection flux in each energy bin, interpolated to a line for clarity) derived from the mean spectra presented in \citet{marinucci14} for SWIFT~J2127.4+5654 and in \citet{walton14} for NGC~1365.  Both sources are well fitted by an absorbed continuum, relativistic reflection and some distant neutral reflection, and we refer the interested reader to those papers for details of the spectral results.  From Fig.~\ref{model} we see that the shapes of the high-frequency lag-energy spectra follow the reflection fractions for each source.  We can use the reflection fractions between two channels of interest to convert the measured lag into an intrinsic lag.  

We start with the case of SWIFT~J2127.4+5654.  The Compton hump (13--30~keV) is measured to lag behind the zero-point continuum (4--5~keV) by 500~s.  The reflection fraction at the Compton hump is around 50\%, and the reflection fraction of the continuum band is around 10\%. Therefore, the measured lag of 500~s is about 40\% of the intrinsic lag, i.e. 1250~s.  The intrinsic reverberation lag is associated with the time it takes for the X-rays to travel from the source, to the disc, and then up to the same height as the source.  For simplicity, if we assume a face on disc for these Seyfert galaxies, then the height of the source above the disc is half of the intrinsic lag, or 625 light seconds, in the case of SWIFT~J2127.4+5654.  We can use black hole mass estimates from the literature to convert this coronal height into gravitational radii (e.g. a 10~s light travel time for a $2 \times 10^6 M_{\odot}$ black hole corresponds to a distance of 1~$r_{\mathrm{g}}$).  The best estimate of the black hole mass in SWIFT J2127.4+5654 is $1.5 \times 10^7 M_{\odot}$ \citep{malizia08}, which puts the corona at $\sim 8~r_{\mathrm{g}}$ above the disc.  This source height is larger than the average source height of 4~$r_{\mathrm{g}}$ found by fitting a sample of lag-frequency spectra \citet{emm14}, but perhaps that is not surprising given that SWIFT~J2127.4+5654 has a measured intermediate spin, while many in that sample are measured to have maximally spinning black holes.

For NGC~1365, the Compton hump (30--50~keV) is measured to lag behind the zero-point continuum (3--4~keV) by $\sim 600$~s.  The reflection fraction is around 60\% at the Compton hump, and 20\% for the continuum band,  which puts the intrinsic lag at 1500~s, and a coronal source height at half that light travel distance, 750 light seconds.  The black hole mass for this source is tentative, and mass estimates have ranged from $2\times 10^6 M_{\odot}$ \citep{kaspi05} to $6\times 10^7 M_{\odot}$ \citep{marconi03}.  The large amplitude and low-frequency of the Fe~K lag support the higher black hole mass estimate for this source.  Using the result from \citet{emm14} that the average source height is $4~r_{\mathrm{g}}$, the black hole mass would then be $\sim 4 \times 10^7 M_{\odot}$.  At these small distances close to the black hole, the Shapiro delay will likely be an important effect, but for this simple calculation, we do not include it here \citep[for more on the lags due to the Shapiro delay, see][]{wilkins13}.  Reverberation lags give a physical distance for the scales around a black hole and therefore could be strong probes of the black hole mass \citep[See][for more on fitting the lag-frequency spectrum with General Relativistic ray tracing models to obtain a consistent estimate of the black hole mass]{emm14}.

\citet{fabian14} recently argued that a spin measurements can only clearly be made when the source height is small, less than around $10~r_{\mathrm{g}}$. Our lag results confirm a small source height, which gives us additional confidence in the spectral line-fitting results.

\subsection{Interpretation of the low-frequency lags}

\subsubsection{The propagation lag in SWIFT~J2127.4+5654}

The large amplitude low-frequency lags have often been observed in variable AGN \citep{papadakis01,mchardy04,arevalo06b,mchardy07}.  The origin of the hard lag is still not well understood. In the prevailing phenomenological model of \citet{kotov01}, which we refer to as the `propagation' model, perturbations are introduced in the accretion flow at a broad range of radii.  These perturbations propagate inwards on the diffusion timescale. This is a multiplicative effect, causing variability at very large radii to correlate with the variability at small radii. These perturbations can modulate the X-ray emitting region, and if the emissivity of the soft emission extends to larger radii than the hard emission, this will naturally cause the hard photons to lag behind the soft.  As this is a multiplicative effect, the perturbations from large radii do {\em not} imply that the X-ray source extends out to those large radii, and the corona can still be contained within a few gravitational radii of the central black hole.

The low-frequency lags have been shown to exhibit a featureless increase with energy in both X-ray binaries \citep{nowak99,kotov01} and in AGN, however, in SWIFT~J2127.4+5654, we find some structure in its lag-energy spectrum at around 7~keV (Fig.~\ref{2127_low_lagen}).  At high-frequencies (Fig.~\ref{2127_high_lagen}) there is a clear dip at ~7~keV, and so it is possible that the dip in the high-frequency lags is corresponds to the increase in the low-frequency lag at 7~keV.  In other words, we are seeing contamination in the low-frequency lags from the high-frequency lags. This has been seen before, in 1H0707-495, where a sharp increase in the low-frequency lag occurs at the same energy (1~keV) as a dip in the high-frequency lag-energy spectrum \citep{kara13a}.  Similar contamination effects at the soft excess are seen in the principal component analysis of Seyfert galaxies \citep{parker14}.  

\subsubsection{A variable absorption model for NGC~1365}

\begin{figure}
\includegraphics[width=\columnwidth]{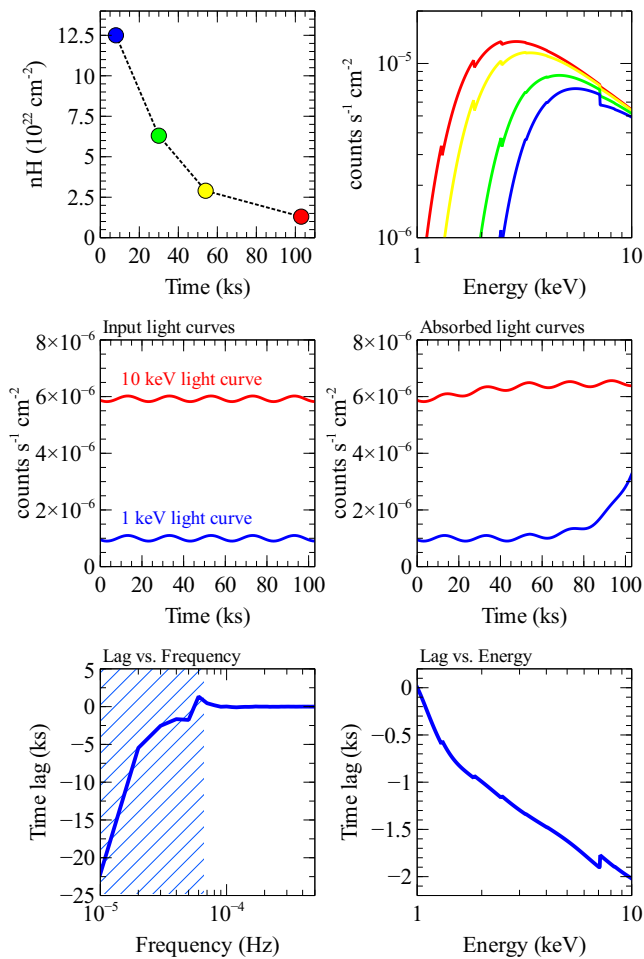}
\caption{A simple model explaining the low-frequency soft lag found in NGC~1365 in the {\em XMM-Newton} band.  The top-left panel shows the data of the $N_{\rm H}$ column density of the neutral absorber decreasing throughout the observation.  The top-right panel shows an absorbed power law with different column densities, corresponding to the $N_{\rm H}$ observed at different times. The middle panels show simple sinusoidal light curves at 1 and 10~keV. The light curves on the left are the input light curves (with no intrinsic lag between them), and the ones on the right are the light curves affected by the changing absorption.  The hard band light curve is affected by the changing absorption before the soft band, which causes a soft lag.  The bottom figures show the lag-frequency spectrum (between 1 and 10~keV) and lag-energy spectrum (for frequencies $[1-7] \times 10^{-4}$~Hz).  The amplitude and shape of the lag-energy spectrum from this simple model are similar to those found in the data in Fig.~\ref{1365_low_lagen}.}
\label{absorbs}
\end{figure}

The third orbit of NGC~1365 shows a low-frequency soft lag (Fig.~\ref{1365_low_lagen}). This is different than the low-frequency lags found in SWIFT~J2127.4+5654 or any other AGN yet studied, where we typically find a low-frequency hard lag. 
We can understand this low-frequency soft lag in terms of a change of column density in the neutral reflector during the observation. Using Principal Component Analysis, \citet{parker14} has recently shown that for NGC~1365, absorption can explain much of the variability below 3~keV at long timescales.  Furthermore, \citet{walton14} showed that in the third orbit (where we see this low-frequency soft lag), the column density is decreasing with time.  Physically, this can be understood as an eclipsing cloud that is moving out of our line of sight, so the nuclear region is becoming less obscured.  The hard photons can travel further through the high column material, and so at first, when the source is obscured, only the hard photons are able to penetrate the cloud. As the cloud moves out of our line of sight, we are able to see also the soft photons. Therefore this low-frequency lag is associated with a {\em change} in the column density.  The amplitude of the lag will depend on how fast the eclipsing cloud is moving (i.e. the rate at which the column density decreases).  

We test this hypothesis through a simple model, shown in Fig.~\ref{absorbs}.  In the top left panel, we show the results from fits to the photon spectrum in \citet{walton14}, where the column density of the neutral absorber systematically decreases during the third orbit.  The top right panel shows what the absorbed spectrum looks like at each of those times.  For simplicity, we have used an absorbed powerlaw with changing column density.  We start with light curves that have zero intrinsic lag between different energy bins.  The middle-left figure shows simple sinusoid light curves at 1~keV and 10~keV. We then evolve the flux of the light curves in time, as the column density decreases.  The middle-right figure shows the resulting light curves, and it is clear that the flux of the 10~keV light curve begins to increase before the 1~keV light curve (because the soft photons are most affected by the neutral absorber).   The bottom-left figure shows the lag vs. frequency between the 1~keV and 10~keV light curves, which shows the soft band lagging at low frequencies.  Finally,e calculate the lag-energy spectrum (bottom-right) at the frequency range $[1-7] \times 10^{-5}$~Hz, just as for the observed low-frequency lag-energy spectrum.  The amplitude of the lag in this simple model is similar to what we find in the data. We note that in this model we do not account for dilution from propagation lags (which likely exist). This will cause the observed amplitude of the lag to decrease. This effect could account for the slight difference that we find between our simple model and the data.  This model predicts a constant lag above 10~keV because the absorption does not affect the higher energies.

The low-frequency soft lag in Orbit 3 is a transient phenomenon, not present in all orbits, which is consistent with the eclipsing cloud interpretation.  Orbit 1 shows low variability power and coherence, but there is a slight negative lag at the highest energies, as in Orbit 3.  The column density varies throughout the observation, though the general trend is a decrease \citep{walton14}. Orbit 2 has a constant column density throughout the observation, and so there would be no lag due to absorption changes in this source.  We find a low-frequency hard lag, similar to other Seyfert galaxies, including SWIFT~J2127.4+5654. The hard lag in the second orbit may be associated with the propagation lag. The column density in Orbit 4 increases throughout the observation, opposite to Orbit 3.  Therefore, we would expect the lag to increase due to the increase in column density in this observation. The data quality is poor in this Orbit, though the lag-energy spectrum does have a positive slope.

\section{Conclusions}
\label{conclusion}

We have presented the lag analysis of the joint {\em XMM-NuSTAR} observations of SWIFT~J2127.4+5654 and NGC~1365. Our main findings are:

\begin{enumerate}
\item SWIFT J2127.4+5654, with an intermediate spin black hole, shows a narrower Fe K lag than sources with maximally spinning black hole, and also shows a clear lag associated with the Compton hump.
\item The amplitude of the iron K lag and the Compton hump lag in SWIFT~J2127.4+5654 are consistent with each other, and can be well described by a light travel time of 1000~s between the corona and the accretion disc.
\item NGC~1365 has a very clear iron K lag in the least absorbed {\em XMM-Newton} observation.  The lag above 10~keV appears to increase at the energy of the Compton hump, though the lags are not very well constrained.
\item At low frequencies, NGC~1365 does not show a featureless hard lag, rather there appears to be a soft lag at the {\em XMM-Newton} band.  This can be understood as the effect of an eclipsing cloud that moves out of our light of sight during the observation, thus decreasing the column density, which causes the hard photons to respond before the soft at these long timescales. 
\end{enumerate}

MCG-5-23-16 \citep{zoghbi14}, SWIFT~J2127.4+5654 and NGC~1365 are the first sources to be analysed for their high-frequency lags in the {\em NuSTAR} band.  There is clear evidence for the Fe K lag and associated Compton hump lag in the high-frequency lag-energy spectra, especially in SWIFT J2127.4+5654. The iron K and now the Compton hump lag measurements are completely independent of spectral modelling, and are strong confirmation of relativistic reflection off an ionised accretion disc.

\section*{Acknowledgements}

We thank the anonymous referee for helpful comments.  EK thanks the Gates Cambridge Scholarship. ACF thanks the Royal Society. EK, ACF, AM and GM acknowledge support from the European Union Seventh Framework Programme (FP7/2007-2013) under grant agreement n.312789, StrongGravity.  AM and GM acknowledge financial support from Italian Space Agency under grant ASI/INAF I/037/12/0-011/13.  This work is based on observations obtained with {\em XMM-Newton}, an ESA science mission with instruments and contributions directly funded by ESA Member States and NASA. This work was supported under NASA Contract No. NNG08FD60C, and made use of data from the {\em NuSTAR} mission, a project led by the California Institute of Technology, managed by the Jet Propulsion Laboratory, and funded by the National Aeronautics and Space Administration. We thank the {\em NuSTAR} Operations, Software and Calibration teams for support with the execution and analysis of these observations. This research has made use of the {\em NuSTAR} Data Analysis Software (NuSTARDAS) jointly developed by the ASI Science Data Center (ASDC, Italy) and the California Institute of Technology (USA).

\appendix

\section*{Appendix A: More on the maximum-likelihood method}

The maximum-likelihood method for measuring time lags in unevenly sampled X-ray light curves was first presented in \citet{zoghbi13}.  In this appendix, we take the reader through an example of calculating the PSD in one energy band in order to demonstrate how the method works.  While this example is for the PSD, the same procedure applies for measuring the time lags; Instead of computing the autocorrelation in one light curve, we compute the cross correlation matrix between two light curves in different energy bands.  The Fourier transform of the cross correlation matrix gives us the cross spectrum, consisting of an amplitude and phase difference.

The basic principal is that we start with one unevenly sampled {\em NuSTAR} light curve, $x$ of length time $T$, with $N$ elements.  We are in the regime where we use uniform time bins of width $\Delta t$, but we are periodically missing data due to the orbital gaps.  While these gaps prohibit us from using the standard Fourier techniques for timescales longer than each orbital period, we can take the autocorrelation of the entire light curve $x$.  We compare this autocorrelation to a model correlation matrix $\mathrm{C}_x$, with dimensions $N \times N$.  Once we have a model correlation matrix that is a good description of the data, we use the relation that the autocorrelation is the Fourier transform of the power spectrum to obtain an estimate for the power spectrum.  

In practice, we start with an initial model of the power spectrum.  We model the power spectrum as a step function, where our model parameters $a_p$ are the average power in each pre-defined frequency bin.  This is the best approach when the intrinsic shape of the PSD or cross spectrum is unknown (which is the case for the cross spectrum).  Then we take the Fourier transform of our initial model PSD to convert it into the model correlation matrix.  The power in our PSD is set to zero below the lowest frequency, $f_{\mathrm{min}}$, so that the integration of the power spectrum converges. We then correct for this bias by adding additional low-frequency power that is estimated from Monte Carlo simulations.  We maximize the model parameters $a_p$ by constructing a likelihood function between the model correlation matrix and our light curve $x$ \citep[Eq.~5 in ][]{zoghbi13}.  The standard is to maximize the log of the likelihood instead of just the likelihood.  The structure of the likelihood function is relatively smooth and converges within a few iterations. Unlike previously stated in \citep{miller10}, the number of iterations {\em is} dependent on the length of the light curve and the number of model parameters (i.e. the number of pre-defined frequency bins).

\begin{figure}
\includegraphics[width=\columnwidth]{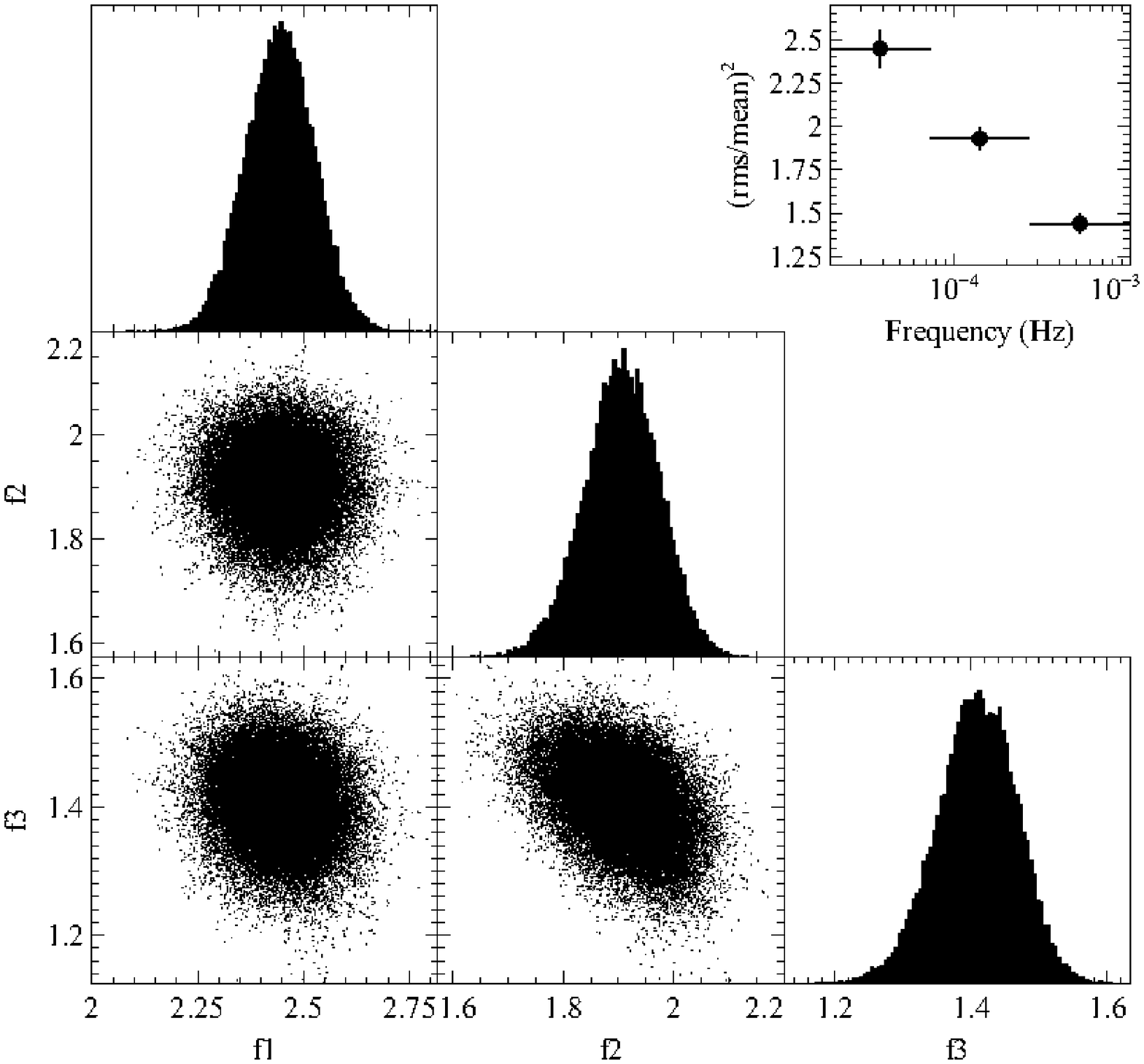}
\caption{The 1D and 2D probability distributions using the MCMC approach for our three model parameters in the case of fitting for the PSD in three pre-defined frequency bins for SWIFT~J2127.5+5654.  The 1D distributions are nearly Gaussian and the 2D distributions show that there is little correlation between broad frequency bands.  The panel on the top-right shows the resulting PSD for our three frequency bins (f1, f2 and f3).}
\label{mcmc}
\end{figure}

We use SWIFT~J2127.4+5654 for this example.  The maximum likelihood is computed from a PSD with 3 equally space logarithmic frequency bins from  $f_{\mathrm{min}} \sim 1/T$ up to $f_{\mathrm{max}}$, around the Nyquist frequency, $f_{\mathrm{Nyq}}=1/(2\Delta t)$, where $\Delta t$ is the width of our time bins (512~s for this case).  We chose the lower frequency limit to be just below $f_{\mathrm{min}}$ to account for red noise leakage due to the finite size of the observation.  This corrects for the bias at low frequencies. At high frequency, the bias is due to aliasing above $f_{\mathrm{Nyq}}$. \citet{babu10} and others note that for unevenly sampled light curves the aliasing effect does not necessarily begin at the Nyquist frequency because $\Delta t$ is not constant. In our case, where our sampling is largely uniform, just with missing data due to the orbital gaps, this effect is small.  Also, as the reported lags are well below the Nyquist frequency, this high-frequency bias is not a concern.

For the PSD of SWIFT~J2127.4+5654 in 3 frequency bins, the likelihood converges in 13 iterations.  We compute the errors by stepping through the parameters and taking the 68\% uncertainty as the value that changes $-2$log($\mathcal{L}/\mathcal{L}_{\mathrm{max}}$) by 1 or similarly, by using Markov Chain Monte Carlo (MCMC) to map the full probability space for each parameter directly (which is computationally more expensive).  For our example, it took 62 minutes on a single Intel(R) Core with 3.4 GHz processor to map the full probability space.  By contrast, it took only 0.6 minutes to step through each parameter and take the value that changes $-2$log($\mathcal{L}/\mathcal{L}_{\mathrm{max}}$) by 1.  This is the quicker method when the number of parameters is less than $\sim 20$. 

Fig.~\ref{mcmc} shows the 1D and 2D probability distributions for each of the three model parameters using the MCMC method.  For cases where the probability distributions are Gaussian and not correlated between parameters (as in our case), the error bars are well estimated in the fitting procedure by the Fisher matrix, the second derivative of the log-likelihood, which basically measures how fast the likelihood function falls around its maximum.  As this does not require further iteration, it is the most computationally efficient error calculation, though it is important to check that the model parameters are not correlated or the Fisher matrix will underestimate the error bar size.

The panel in the top-right of Fig.~\ref{mcmc} shows our final PSD for SWIFT~J2127.4+5654.  The same procedure is used for computing the cross spectrum for the time lags that are presented in this paper.

\renewcommand{\thefigure}{\Alph{figure}}

\section*{Appendix B: The choice of reference band}

In Fig.~\ref{model}, we overplot the {\em XMM-Newton} and {\em NuSTAR} lags for SWIFT~J2127.4+5654 and NGC~1365.  The reference band used to calculate the lags from these two datasets are different (0.3--10~keV for {\em XMM-Newton} and 3--50~keV for {\em NuSTAR}).  We show here, that this only results in a change in the normalization of the lag, and so we can compare the two by simply scaling the lag-energy spectra by some constant.  The relative lag between energy bins is the relevant value, and this remains the same for any coherent reference band.  This result has been tested through simulations, and discussed in \citet{zoghbi11}, and we demonstrate this point further here. 

\begin{figure}
\includegraphics[width=\columnwidth]{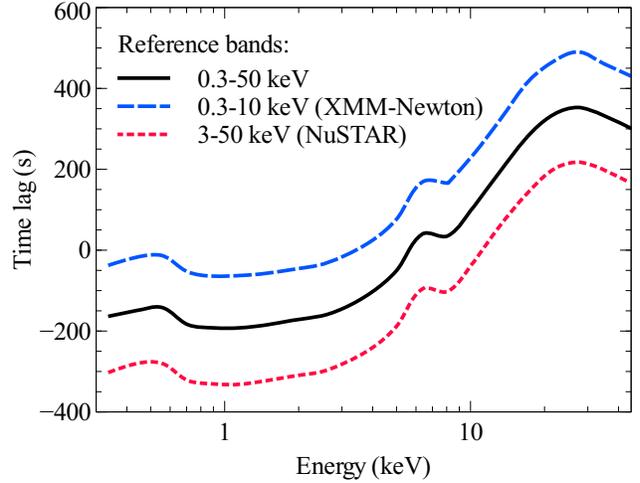}
\caption{A simple demonstration of the lag-energy spectrum for SWIFT~J2127.4+5654 using the entire band from 0.3--50~keV in black, the {\em XMM-Newton} band from 0.3--10~keV in blue, and the {\em NuSTAR} band from 3--50~keV in red.  We show that the shape of the lag-energy spectrum does not change, but the normalization does, so we can directly compare the lags from the two instruments by simply scaling the absolute lag.}
\label{ref}
\end{figure}

We construct a simple demonstration in Fig.~\ref{ref} for the case of SWIFT~J2127.4+5654.  The lag-energy spectra shown here are the result of interpolation between twenty equally spaced logarithmic bins from 0.3--50~keV.  We computed 1000 Monte Carlo light curves in each bin using the method of \citet{timmer95}. To reduce scatter for the figure, the light curves do not include Poisson noise, but the overall result is the same including Poisson noise.  Each light curve contains some primary emission and some reflected emission, imposed by the fraction of each component in the mean spectrum from \citet{marinucci14}.  The reflected emission was delayed by 1250~s (the value calculated for the intrinsic lag in Section~\ref{highfreq}).  We assume unity coherence between the primary and reflected components, as suggested by the high level of coherence in the data.  If there is some additional variable, but non-coherent component in the reference bands, this will cause the shapes of the lag-energy spectra to be different, but we find no evidence of that in these sources (e.g. the right panel of Fig.~\ref{xmm_lag_freq}).  

Fig.~\ref{ref} shows what the lag-energy spectra look like for different reference bands.  The solid black line shows the lag-energy spectrum using the entire 0.3--50~keV band, and the blue dashed and red dotted lines show the lag-energy spectra for the {\em XMM-Newton} and {\em NuSTAR} reference bands, respectively.  The shape of the lag-energy spectrum does not change (it follows the reflection fraction, as shown in Fig.~\ref{model}), yet the normalization is different.  The difference between the lag-energy spectra with {\em XMM-Newton} and {\em NuSTAR} is $\sim 270$~s.  The observed difference between the {\em XMM-Newton} and {\em NuSTAR} lag-energy spectra is also 270~s.  This gives us confidence that we are observing the same features in the lags when comparing the two datasets.

\label{lastpage}

\end{document}